\documentclass{aa}
\usepackage{graphicx}
\usepackage{txfonts}
\begin{document}

\title{Search for rotating radio transients in monitoring data\\ for three years}
   
   \author{S.A. Tyul'bashev\inst{1}
          \and
           M.A. Kitaeva\inst{1}
           \and
           D.V. Pervoukhin\inst{2}
          \and
          G.E. Tyul'basheva\inst{3}
          \and
          E.A. Brylyakova\inst{1}
          \and
          \\
          A.V. Chernosov\inst{4}
          \and
          I.L. Ovchinnikov\inst{5}
          }

  \institute{Lebedev Physical Institute, Astro Space Center, Pushchino Radio Astronomy Observatory\\
142290, Radioteleskopnaya 1a, Moscow reg., Pushchino, Russia\\
  \email{serg@prao.ru}
                     \and
    LLC Impact Electronics, 127055, Novoslobodskaya str. 14/19 building 8, Moscow, Russia
                     \and
 Institute of Mathematical Problems of Biology, brunch of Keldysh Institute of Applied Mathematics\\
142290, Vitkevich 1, Moscow reg., Pushchino, Russia
                     \and
FGBOU VO Moscow State University of Psychology and Education, 127051, Sretenka str. 29, Moscow, Russia
                     \and
Skobeltsyn Institute of Nuclear Physics, Lomonosov Moscow State University, 119234, Moscow,  Russia
             }

   \date{Received ; accepted}
   
\abstract
{The search for rotating radio transients (RRAT) was done at a frequency of 111 MHz, in daily observations carried out on the radio telescope, a Large Phased Array (LPA) at declinations $-9^o< \delta < +42^o$. 19 new RRATs were discovered for dispersion measures (DM) from 2.5 to 72.6~pc cm$^{-3}$. Estimates of the periods were obtained for three RRATs. Two of them (J0408+28; J0440+35) are located at distances of 134 and 136 pc from Sun and are among the closest of all known RRATs.}


   \maketitle
%

\section{Introduction}

Rotating radio transients (RRATs) were discovered in 2006 as dispersed pulses in archived data from the 64-meter telescope in the Parkes (\citeauthor{McLaughlin2006}, \citeyear{McLaughlin2006}). Unlike conventional pulsars, which emit a pulse at each or almost every revolution of a neutron star, RRATs do not emit pulses regularly, and hours can pass between two recorded pulses (\citeauthor{}, \citeyear{})\citep{McLaughlin2006}. 

The search for RRAT pulses (\citeauthor{Cordes2003}, \citeyear{Cordes2003}; \citeauthor{Fridman2010}, \citeyear{Fridman2010}; \citeauthor{Swinbank2015}, \citeyear{Swinbank2015}; \citeauthor{Zhang2020}, \citeyear{Zhang2020}) and the search for pulsars (\citeauthor{Burns1969}, \citeyear{Burns1969}; \citeauthor{Staelin1969}, \citeyear{Staelin1969}; \citeauthor{Hankins1975}, \citeyear{Hankins1975}; \citeauthor{Eatough2010}, \citeyear{Eatough2010}) are conducted in different ways.  Unlike conventional pulsars, in which it is possible to accumulate a signal by adding data with a known period over long time intervals, for the search for dispersed pulses, it is necessary to have a high instantaneous sensitivity. That is, the search for pulsars can be carried out when their individual pulses are not visible in the raw data after compensation for the dispersion measure (DM). To search for RRAT, it is necessary that the ratio of the signal (pulse height; $A$) to noise (root mean square deviation - RMS - $\sigma_n$  in the noise track) be greater than 6-7: S/N=$A/\sigma_n > 6-7$. In addition, due to the random appearance of RRAT pulses and unpredictable waiting time, long-term observations of each point in the sky are needed to find them.

There is no generally accepted definition of RRAT. In this paper, we adhere to the definition of RRAT as a special group of pulsars that are detected by individual irregular pulses (\citeauthor{Burke-Spolaor2010}, \citeyear{Burke-Spolaor2010}). It is assumed that RRAT can be ordinary pulsars with individual strong pulses (\citeauthor{Zhou2023}, \citeyear{Zhou2023}), pulsars with a very wide distribution of pulses by energy (\citeauthor{Weltevrede2006}, \citeyear{Weltevrede2006}), pulsars with giant pulses (\citeauthor{Brylyakova2021}, \citeyear{Brylyakova2021}; \citeauthor{Tyulbashev2021}, \citeyear{Tyulbashev2021}), pulsars with nulling (\citeauthor{Zhang2007}, \citeyear{Zhang2007}; \citeauthor{Wang2007}, \citeyear{Wang2007}).

To date, several hundred RRATs can be found in catalogs\footnote{http://www.atnf.csiro.au/people/pulsar/psrcat/}$^,$\footnote{http://astro.phys.wvu.edu/rratalog/}$^,$\footnote{https://bsa-analytics.prao.ru/transients/rrat/}$^,$\footnote{http://zmtt.bao.ac.cn/GPPS/} (\citeauthor{Manchester2005}, \citeyear{Manchester2005}; \citeauthor{Han2021}, \citeyear{Han2021}), but there is no generally accepted opinion about the nature of RRAT yet. Are they a special selection of pulsars (a new type), or are RRATs ordinary pulsars included in long-known samples? Perhaps these different samples are not related to each other, and observers see only a general manifestation of neutron star activity in the form of rarely appearing dispersed pulses.

Telescopes that currently have the highest sensitivity in their ranges are used to search for RRATs. In the meter wavelength range, RRATs were found or investigated on three antenna arrays. These are Low Frequency Array (LOFAR; \citeauthor{vanHaarlem2013} (\citeyear{vanHaarlem2013}), Murchison Widefield Array (MWA; \citeauthor{Tingay2013} (\citeyear{Tingay2013}), Large Phased Array (LPA; \citeauthor{Shishov2016} (\citeyear{Shishov2016})). The search for RRATs on LOFAR was carried out on recordings, the duration of which was 1 hour for each direction in the sky (\citeauthor{Sanidas2019}, \citeyear{Sanidas2019}). The search for RRATs on MWA with duration 1.33 hour for each direction is declared in the paper \citeauthor{Bhat2023} (\citeyear{Bhat2023}), but no new RRATs have been discovered yet. The RRAT search on the LPA was performed several times. Most of the RRATs were discovered when processing semi-annual round-the-clock observations (\citeauthor{Tyulbashev2018b}, \citeyear{Tyulbashev2018b}). For 6 months, the accumulation at each point in the sky amounted to about ten hours. It was possible to discover 25 new RRATs and the pulses of almost a hundred known pulsars\footnote{https://bsa-analytics.prao.ru/transients/pulsars/}. Since the time intervals between consecutive pulses of previously detected RRATs can be several hours (\citeauthor{McLaughlin2006}, \citeyear{McLaughlin2006}), we expected that observations within ten hours would allow us to detect all or almost all of the RRATs available for observations on the LPA. However, subsequent studies (\citeauthor{Logvinenko2020}, \citeyear{Logvinenko2020}; \citeauthor{Tyulbashev2022a}, \citeyear{Tyulbashev2022a}) have shown that there are RRATs in which tens of hours can pass between consecutive detections. 

The main purpose of this study is to search for new RRATs in the three-year observation interval on the LPA radio telescope. The accumulation at each point in the sky is approximately 65 hours. The daily survey covers 17,000 sq.deg.

\section{Observations and data processing}


The observations were carried out on the LPA radio telescope of Lebedev Physical Institute (LPI) (\citeauthor{Shishov2016}, \citeyear{Shishov2016}; \citeauthor{Tyulbashev2016}, \citeyear{Tyulbashev2016}). LPA is a full-power antenna array consisting of 16,384 dipoles. Several radio telescopes have been created on the basis of an antenna field measuring approximately $200 \times 400$~m. In the project  Pushchino multibeams pulsar search (PUMPS; \citeauthor{Tyulbashev2022b} (\citeyear{Tyulbashev2022b})) LPA3 radio telescope is used, which has 128 stationary beams. The beams are located in the plane of the meridian. They overlap at the 0.405 level and cover declinations $-9^o< \delta < +55^o$. The dimensions of the receiving beam are about $0.5^o \times 1^o$. The passage of a source through the meridian occurs once a day (one observation session per day) and takes about 3.5 time minutes at half power. The instantaneous viewing area is about 50 sq.deg. The observations are carried out around the clock on the central frequency of 110.25 MHz, in the 2.5 MHz band. The band is divided into 32 frequency channels with the width of 78 kHz. The sampling of a point is 12.5 ms. At the end of 2014, observations were started in 96 beams cover declinations $-9^o< \delta < +42^o$, the remaining 32 beams were connected to the recorders in 2021-2022.

There are several standard steps in a regular RRAT search: subtracting the baseline; sorting through different DMs; estimation of the RMS deviations of noise; search for pulses having S/N greater than the set value. During the search, the data is cut into pieces, in each of which the presence of a pulse is checked \cite{Tyulbashev2018b}. 

To estimate the peak flux density ($S_{p}$) of pulses, well-known antenna parameters were used: the effective area depending on the height of the source above the antenna, the position of the source relative to the center of the beam, the frequency band, the sampling time, the background temperature in the direction of the transient. The background temperature was recalculated from a frequency of 178 MHz (see maps in \citeauthor{Turtle1962} (\citeyear{Turtle1962})) to a frequency of 111 MHz based on the dependence $T_b\sim \nu^{-2.55}$.

To obtain the result in this work, we processed data for the period August 2014 - December 2017 obtained on declinations $-9^o< \delta < +42^o$, and checked $\sim 10^{12}$ of chopped pieces of raw data. The search was conducted for transients with DM<100 pc cm$^{-3}$. With such a large number of sliced pieces of data, we can detect a few thousands (S/N=6) false transients. However, the probability of re-detecting a transient with the same coordinate and on the same DM is negligible. In this paper, we have chosen the detection criterion for reliable discovering of three pulses having S/N>6.

The processing program has selected $4.5 \times 10^6$ RRAT candidates. A strong pulse in the proposed treatment can be detected on many DMs, but knowing its exact time coordinate, you can leave the candidate for RRAT only for the DM where it shows the maximum S/N. To reduce the amount of interferences, we have done additional filtering of candidates using a recurrent neural network (RNN) using LSTM layers (\citeauthor{Hochreiter1997}, \citeyear{Hochreiter1997}). As a result, we had $\sim 10^6$ RRAT candidates remaining. The vast majority of the remaining candidates are pulses of known pulsars observed in the main, side and back lobes of LPA3. Their coordinates and DMs are known, because previously the RRAT search was conducted on a semi-annual interval (\citeauthor{Tyulbashev2018b}, \citeyear{Tyulbashev2018b}). Tens of thousands of pulses are observed for some pulsars. We have excluded from the search the directions in which pulses from known pulsars are visible. Such pulses account for approximately 98\% of all remaining pulses for verification. Thus, approximately $10^6 \times 0.02=20,000$ pulses were visually checked.  The analysis showed that most of the detected 20,000 pulses also belong to known pulsars observed in the side and back lobes of LPA3, or are an interference. 

\begin{figure*}
   \centering
 \includegraphics[width=0.95\textwidth]{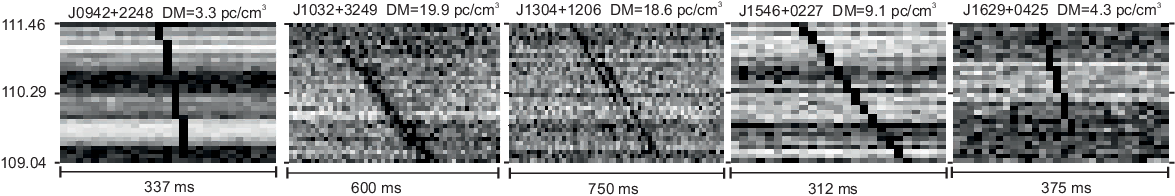}
   \caption{Dynamic spectra of some RRAT candidates, for which 1-2 pulses were detected at an interval of 9 years.}
    \label{fig:fig1}%
    \end{figure*}

 \begin{table*}
	\centering
	\caption{Characteristics of the found RRATs}
	\label{tab:table1}
	\begin{tabular}{lccccccccc} 
		\hline
name &   $\alpha_{2000}$&  $\delta_{2000}$&$P$(s)& DM(pc cm$^{-3}$)&  S/N  &$W_{0.5}$(ms)&$S_p$(Jy)&  n & D (pc) \\
	\hline	
J0313+36&$03^h13^m13^s$&  $+36^o15^\prime$&      &   20.8$\pm$1.5&     12.1&    53 &       4.1&  3 & $1041^{+68}_{-24}$\\ 
J0408+28&     04 08 34&      +28 45&       2.92 &   2.5$\pm$1.0 &     23.5&    13 &       8.0&   6 & $134^{+5}_{-11}$\\ 
J0440+35&     04 40 00&      +35 22&       2.23 &   2.6$\pm$1.0 &     48.2&    21 &       14 &   6 & $136^{+5}_{-10}$\\
J0601+38&     06 01 02&      +38 46&            &   20.9$\pm$1.5&      6.3&    15 &       1.9&   3 & $265^{+90}_{-71}$\\ 
J0630+23&     06 30 43&      +23 13&            &   12.4$\pm$1.0&     73.3&    40 &       22 &   3 & $160^{+1}_{-1}$\\
J1104+14&     11 04 18&      +14 13&            &   23.2$\pm$1.5&     15.1&    23 &       13 &   3 & $25000^{+0}_{-21000}$\\ 
J1157+25&     11 57 44&      +25 40&            &   8.85$\pm$1.0&     25.3&    15 &       8.1&   6 & $848^{+114}_{-105}$\\ 
J1404+14&     14 04 39&      +14 48&            &   13.3$\pm$1.5&     11.4&    16 &       3.7&   3 & $1228^{+244}_{-158}$\\   
J1530+00&     15 30 42&      +00 47&            &   13.4$\pm$1.5&     11.6&    23 &       5.8&   14 & $929^{+118}_{-108}$\\
J1605-07&     16 05 12&      -07 37&       1.81 &   4.8$\pm$1.0 &     64.4&    25 &       54 &   20 & $294^{+61}_{-63}$\\
J1826-08&     18 26 00&      -08 37&            &   19.9$\pm$1.5&     7.2 &    21 &       20 &   8  & $372^{+74}_{-84}$\\
J1830+18&    18 30 15 &      +18 15&            &   57.6$\pm$2.0&     15.5&    35 &       10 &   45 & $3471^{+219}_{-206}$\\
J1943+09&    19 43 45 &      +09 40&            &   46.0$\pm$2.0&     13.4&    26 &       9.0&   10 & $1819^{+56}_{-55}$\\
J2007+13&    20 07 06 &      +13 00&            &   67.4$\pm$2.0&     11.8&    48 &       4.9&   3  & $3795^{+166}_{-159}$\\
J2019-07&    20 19 44&      -07 45&            &   24.7$\pm$1.5&     20.4&    90 &       17 &   4 & $1360^{+90}_{-81}$\\
J2119+40&    21 19 19 &      +40 51&            &   72.6$\pm$3.0&     8.8 &    42 &       3.3&   4  & $3913^{+94}_{-99}$\\
J2251+14&    22 51 04 &      +14 00&            &   10.2$\pm$1.5&     28.4&    37 &       24 &   7  & $769^{+133}_{-125}$\\
J2337-04&     23 37 03&      -04 40&            &   15.3$\pm$1.5&     16.1&    16 &       10 &   3  & $1558^{+266}_{-224}$\\
J2359+06&    23 59 09 &      +06 32&            &   19.8$\pm$1.5&     43.4&    21 &       17 &   22 & $2163^{+392}_{-302}$\\
		\hline
	\end{tabular}
\end{table*}

\section{Results}

After initial screening, 104 candidates were selected for further research. From 1 to more than ten pulses were recorded for the selected candidates over a three-year interval. For all candidates, the DM was specified according to the strongest pulses, and an additional search for new pulses was carried out using monitoring data recorded from August 2014 to August 2023. Taking into account 3.5 minutes of observations per day for each direction in the sky, approximately 190 hours have been accumulated over 9 years for each RRAT candidate. Therefore, even if the transient emits a pulse once every two days, we must register 3-4 pulses from it. In this paper, we have adopted the criterion of reliable detection of RRAT as three detected pulses.


Out of 104 sources of pulsed dispersed radiation, 19 turned out to be new RRATs. The dynamic spectra and pulse profiles of these RRATs have no special features and are posted on our website\footnote{https://bsa-analytics.prao.ru/transients/rrat/}. In Table~\ref{tab:table1}, columns 1-3 show the name of the transient, coordinates for right ascension and declination for the year 2000. The accuracy of coordinates in right ascension is $\pm 1.5^m$. The accuracy of the declination coordinates is defined as half the distance to neighboring LPA3 beams, located in beams with declinations above and below the beam where the pulse is detected, and is approximately $\pm 15^\prime$. Pulsars J1530+00 and J1830+18 are available on the AO327 (\citeauthor{Deneva2013}, \citeyear{Deneva2013}) and CHIME/FRB (\citeauthor{Dong2023}, \citeyear{Dong2023}) websites\footnote{http://ao327.nanograv.org/}$^,$\footnote{https://www.chime-frb.ca/galactic} under the names J1532+00 and J1830+17, but have not been officially published. Apparently, we independently found theirs pulses. Columns 4-8 show the period ($P$), if it was possible be determine it, DM and the error in determining DM, the observed value of S/N of the strongest pulse, its visible half-width ($W_{0.5}$ – is the width of the pulse at half of its height) and $S_p$. Column 9 shows the number of pulses (n; S/N>6) found over an interval of 9 years.  Column 10 shows the distance (D) to the RRAT. An estimate of the distance to the discovered RRATs can be made using the calculator located on the ATNF website\footnote{https://www.atnf.csiro.au/research/pulsar/ymw16/} and using the YMW16 (\citeauthor{Yao2017}, \citeyear{Yao2017}) model  for calculating. The accuracy of the DM determination depends on the observed S/N of a pulse and on a value of DM. The higher the DM, the greater the detection error due to pulse broadening associated with dispersion smoothing inside the frequency channel and due to scattering (\citeauthor{Kuzmin2007}, \citeyear{Kuzmin2007}). $S_p$ is determined with a large error, since the pulse coordinate can be shifted relative to the center of the receiving beam pattern, both in right ascension and declination. This displacement of the coordinate relative to the center of the antenna beam is unknown, so we cannot make an adjustment that takes into account the transient hitting the edge of the receiving beam. The actual $S_p$ can be up to 1.5-2 times higher than that indicated in column 8. For open transients, a cross-search with the ATNF catalog, as well as with the RRAT search papers for 2023-2024, was done.

Let us note 3 sources that are not included in Table ~\ref{tab:table1}:
\textbf{J1556+01}. This is the RRAT J1555+0108 we found earlier (\citeauthor{Tyulbashev2018a}, \citeyear{Tyulbashev2018a}). In the new search, there are several sessions lasting several minutes, when 2-3 pulses are observed. This allows us to make the period estimate as 0.577 s;\\
\textbf{J1953+30, J2333+20}. 6 and 19 pulses were found for these sources. Their DM and coordinates coincide with the pulsars J1953+30 and J2333+20 previously discovered in PUMPS (\citeauthor{Tyulbashev2017}, \citeyear{Tyulbashev2017}; \citeauthor{Tyulbashev2022b}, \citeyear{Tyulbashev2022b}). Apparently, we detected individual pulses of these pulsars.

17 sources look like RRATs. We were unable to show that the pulses found could be a manifestation of the known strong pulsars observed in the side and back lobes of LPA3. 1-2 pulses are recorded for these RRAT candidates at an interval of 9 years. The existence of transients with 1-2 pulses found in the three-year observation interval probably indicates the existence of new transients in the available nine-year observation interval. We plan to carry out this processing.

\begin{figure}
   \centering
 \includegraphics[width=0.75\columnwidth]{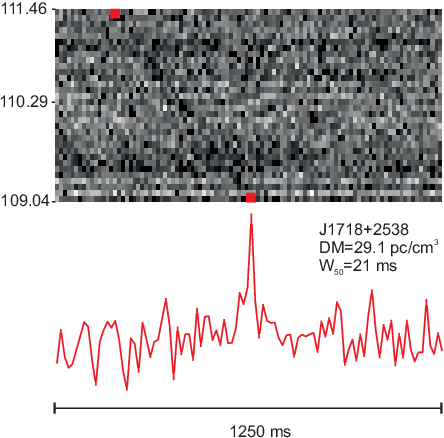}
   \caption{The pulse profile (bottom panel) and dynamic spectrum (upper panel) of J1718+2538. 6 pulses were detected in the transient. The two red dots on the dynamic spectrum are the expected beginning and end of the line along which the dispersion delay line should be located.}
    \label{fig:fig2}%
    \end{figure} 

Since in this work the formal criterion for detecting at least 3 pulses to confirm a RRAT was adopted, they were not included in Table~\ref{tab:table1}. In Figure~\ref{fig:fig1}, we present the dynamic spectra of the strongest RRAT candidates, and we will briefly discuss them in the next section. Light bands are clearly visible on some dynamic spectra. Their appearance is related to the way dynamic spectra are presented.  The weakest signals on the spectra are drawn in white, the strongest are drawn in black. In places where light bands are visible on the dynamic spectra, the signals forming the dispersion delay line were the strongest. 

For 21 sources on dynamic spectra, the dispersion delay line is invisible or poorly visible, while the pulse profile is well visible (S/N>6) (see Figure~\ref{fig:fig2}). Some of these sources have 3 or more pulses. Since these RRAT candidates did not pass visual verification, they were not included in Table~\ref{tab:table1}. On the one hand, the detection of several pulses with close coordinates of right ascension and declination for different dates, as well as with close DM, indicates that the probability of detecting noise signals is low. At the same time, the absence of a pronounced dispersion delay line indicates the likely interference nature of the pulses. For an unambiguous answer to the question whether the detected dispersed signals are new RRATs, or interference, additional search criteria are needed, or an independent test on telescopes with sensitivity higher than LPA3 sensitivity. 13 sources are known pulsars visible in the side lobes of LPA3. For 34 sources, it was possible to show that they are interference.

Table~\ref{tab:table1} contains 19 new RRATs, but the number of detected transients may grow significantly. 17 RRAT candidates  having 1-2 pulses, and some candidates with more than 3 pulses and a poorly distinguishable dispersion delay line, may increase the number of RRATs found with further research up to $\sim$~45-50.

\section{Discussion of the results}

1) The nature of RRAT.\\
As mentioned in the Introduction, the main hypotheses about the nature of RRAT state that rotating radio transients can be part of pulsar samples with a wide distribution of pulses by energy (\citeauthor{Weltevrede2006}, \citeyear{Weltevrede2006}), with giant pulses (\citeauthor{Brylyakova2021}, \citeyear{Brylyakova2021}; \citeauthor{Tyulbashev2021}, \citeyear{Tyulbashev2021}), with sporadic strong pulses (\citeauthor{Zhou2023}, \citeyear{Zhou2023}), with nulling (\citeauthor{Zhang2007}, \citeyear{Zhang2007}; \citeauthor{Wang2007}, \citeyear{Wang2007}). 

Thus, in the paper \citeauthor{Zhou2023} (\citeyear{Zhou2023}) on the RRAT search on the FAST telescope, it was shown that 43 out of the 48 ($\sim$~90\%) transients previously detected in the decimeter wavelength range are ordinary pulsars with sporadic strong pulses. For RRATs discovered on FAST, a regular radiation was detected in only for 24 out of 76 ($\sim$~31.6\%) transients. A smaller percentage of detected pulsars with regular radiation may indicate a lack of FAST sensitivity, and it is likely that regular radiation will be detected with prolonged data accumulation. 

In the search for slow ($P\sim 1$s) pulsars in the meter wavelength range on the LPA3 radio telescope, using power spectra summarized over an interval of several years, it was possible to detect regular weak pulsar radiation in 8 previously discovered RRATs (\citeauthor{Tyulbashev2024}, \citeyear{Tyulbashev2024}). RRATs search on declinations $+56^o < \delta < +87^o$ using the LPA1 radio telescope, it was shown that for known pulsars, peak flux densities in average profiles and in individual pulses in one observation session can differ from tens to hundreds of times (\citeauthor{Tyulbashev2021a}, \citeyear{Tyulbashev2021a}). 

Let us note that pulsars with a wide energy distribution of pulses, pulsars with sporadically appearing strong pulses, and pulsars with giant pulses  will seem similar in search. All these types of pulsars have randomly appearing strong pulses and have regular pulsar radiation. In order to determine which type the transient belongs to, additional research is needed. For example, a power-law energy distribution of pulses may signal the detection of pulsars with giant pulses (\citeauthor{Brylyakova2021}, \citeyear{Brylyakova2021}).

Despite the fact that 90\% of the known RRATs were observed on FAST near Galactic plane as pulsars with regular radiation, it cannot be claimed that all RRATs, when very high sensitivity is achieved, will be detected as ordinary slow pulsars. Thus, in LPA3 observations, the sensitivity in the search for pulsars using summed power spectra can reach 0.1-0.2 mJy (\citeauthor{Tyulbashev2022b}, \citeyear{Tyulbashev2022b}) and at the same time, regular pulsar radiation was detected for only 10\% of new RRATs (\citeauthor{Tyulbashev2024}, \citeyear{Tyulbashev2024}). There are also RRATs in which pulsar radiation is not detected in the summed power spectra, but regular radiation is recorded in individual sessions lasting 3.5 minutes (\citeauthor{Tyulbashev2023}, \citeyear{Tyulbashev2023}).  The peak pulse flux density of many RRATs exceeds tens of Jy at a frequency of 111 MHz (\citeauthor{Tyulbashev2018b}, \citeyear{Tyulbashev2018b}; \citeauthor{Tyulbashev2023}, \citeyear{Tyulbashev2023}) and this paper. Assuming that regular radiation can be detected for all RRATs during accumulation, the ratio between the peak flux densities of the average profiles and the recorded individual pulses should be more than several thousand.

Therefore, for some RRATs, the hypothesis that they can be pulsars with large nulling fraction looks more natural. So, in the paper already mentioned \citeauthor{Zhou2023} (\citeyear{Zhou2023}), it is said that in FAST observations at a central frequency of 1.25 GHz and a frequency band of 500 MHz, regular radiation could not be detected for 10\% of known RRATs detected in GPPS survey. Since standard sessions on FAST last 5 minutes, we can talk about a nulling fraction reaching 99\%. In the paper \citeauthor{Tyulbashev2023} (\citeyear{Tyulbashev2023}), based on observations at 111 MHz, it is said that RRAT J1312+39 was detected, in which only 3 pulses with peak flux densities from 33 to 165 Jy were detected in daily sessions of 3.5 minutes over an interval of 8 years. These flux densities are many times higher than the pulse detection limit at LPA3, which is about 2 Jy for S/N=7 (\citeauthor{Tyulbashev2018a}, \citeyear{Tyulbashev2018a}). For J1312+39, a nulling fraction can reach 99.999\%. Here and further, it is assumed that all RRAT have the same period $P=1$s.

The previous section provides examples of sources (see Figure~\ref{fig:fig1}) with 1-2 pulses registered. These pulses are no different from the pulses of previously detected by RRATs. If they belong to the new RRATs, this means that it is necessary to recognize the existence of transients emitting one pulse per 100-200 hours of observations. The nulling fraction of these pulsars can reach 99.9999\%.

The analysis of the properties of RRATs carried out in the paper \citeauthor{Abhishek2022} (\citeyear{Abhishek2022}) shows that transients are not statistically related to nulling pulsars and are not pulsars at a late stage of their life. Observations of known RRATs on FAST with a sensitivity up to 10 times higher than in the works with the discovery of these RRATs confirm that pulsars with nulling are not the main part of  RRATs (\citeauthor{Zhou2023}, \citeyear{Zhou2023}). For a small number of RRATs in which it was possible to determine the $P$ and $\dot P$ periods, it is shown that, on average, $P$ and $\dot P$ have higher values than for ordinary pulsars (see Fig.10 in \citeauthor{Cui2017} (\citeyear{Cui2017})). Usually, long periods are associated with the age of a pulsar and with its approach to the death line on the $P/\dot P$ plane (\citeauthor{Chen1993}, \citeyear{Chen1993}). However, for RRATs, the points on the $P/\dot P$ plane are generally not pushed against the line of death. This supports the assumption that the observed properties of RRATs are related to some internal processes in the magnetosphere of the neutron star itself (\citeauthor{Abhishek2022}, \citeyear{Abhishek2022}).\\
2) Detection of close transients\\
Out of the 19 transients in Table~\ref{tab:table1}, 3 RRATs are located at a distance of less than 200 pc. These are transients J0408+28 (134 pc), J0440+35 (136 pc), J0630+23 (160 pc). A search in the ATNF catalog for February 2024 shows 22 equally close pulsars, of which 3 are located in the northern hemisphere. 

If the peak flux density in the average profile and the distance to a pulsar are known, its pseudo-luminosity can be estimated ($L=S \times d^2$, where $S$ is the average pulsar flux density in mJy, and $d$ is the distance to the pulsar in kpc; \citeauthor{Lorimer2012} (\citeyear{Lorimer2012})). The usual definition of the average flux density when calculating pseudo-luminosity for transients is not applicable. An insignificant part of the pulses is detected for them, and with a formal approach, when all the periods in the observation interval are summed up, the determined average flux density will be close to zero. However, it is possible to estimate the flux density at one period when a pulse is observed.  Assuming that the observed pulses have a triangular shape and the pulse parameters are known (values of  $P$, $W_{0.5}$, $S_p$ from Table~\ref{tab:table1}), we have estimated $S_{111}= 35.6$ and 131.8 mJy, $L_{111}$=0.64 and 2.44 mJy kpc$^{2}$ for J0408+28 and J0440+35, accordingly.

The ATNF catalog shows pseudo-luminosities for about 600 pulsars at 400 MHz. For a convenience of readers, we will make estimates of pseudo-luminosity at a frequency of 400 MHz using the obtained flux density at 111 MHz. We convert the flux density into 400 MHz, assuming that it is a power-law spectrum and a power-law exponent $\alpha = 1.8 (S \sim \nu^{-\alpha}$) (\citeauthor{Maron2000}, \citeyear{Maron2000}). Then $S_{400}=3.54$ and 13.12 mJy, and $L_{400}=0.064$ and 0.244~mJy kpc$^{2}$  for J0408+28 and J0440+35. There are only 3 pulsars in the ATNF catalog (0.5\% of those having a pseudo-luminosity evaluation) having comparable values of $L_{400}$. These are pulsars J0307+7443, J0613+3731, B1014-53, located at distances from 116 to 386 pc and having pseudo-luminosities from 0.04 to 0.06 mJy kpc$^{2}$. Thus, J0408+28 are included in a short list of pulsars closest to Sun with minimal pseudo-luminosities.

\section{Conclusion}

1)	When checking 104 candidates for rotating radio transients discovered in daily observations lasting 3 years, 19 new RRATs were discovered. 13 RRAT candidates are known pulsars observed in the side and back lobes of LPA3. 17 sources are very similar to conventional RRATs, but for them only 1-2 pulses were detected in daily observations lasting 9 years. 21 sources failed visual check, and their nature is unknown, 34 candidates were associated with interference. Taking into account previously discovered RRATs\footnote{https://bsa-analytics.prao.ru/transients/rrat/} the total number of rotating radio transients discovered at PUMPS has reached $\simeq$ 80.\\ 
2)	For some of the candidates, only one pulse was detected over a time interval equivalent to continuous observations of more than fifty, and possibly more than a hundred hours. If these candidates belong to a sample of pulsars with nulling, then the share of the nulling fraction can  reach 99.9999\%. \\
3)	Distances to 3 discovered RRATs (J0408+28; J0440+35; J0630+23) are less then 200 pc, which indicates the location of the sources in the immediate vicinity of the Sun. 

\section{Acknowledgment}
The study was carried out at the expense of a grant Russian Science Foundation 22-12-00236\footnote{https://rscf.ru/project/22-12-00236/}. The authors thank L.B. Potapova for her help in execution of the paper. The authors thank the anonymous referee for the comments that made it possible to improve the readability of the paper.

\section{Data Availability}
The PUMPS survey not finished yet. The raw data underlying this paper will be shared on reasonable request to the corresponding author. 

\bibliographystyle{aa} 

\begin{thebibliography}{44}
	\expandafter\ifx\csname natexlab\endcsname\relax\def\natexlab#1{#1}\fi
	
	\bibitem[{{Abhishek} {et~al.}(2022){Abhishek}, {Tanushree}, {Hegde}, \&
		{Konar}}]{Abhishek2022}
	{Abhishek}, Malusare, N., {Tanushree}, N., {Hegde}, G., \& {Konar}, S. 2022,
	Journal of Astrophysics and Astronomy, 43, 75
	
	\bibitem[{{Bhat} {et~al.}(2023){Bhat}, {Swainston}, {McSweeney}, {Xue},
		{Meyers}, {Kudale}, {Dai}, {Tremblay}, {van Straten}, {Shannon}, {Smith},
		{Sokolowski}, {Ord}, {Sleap}, {Williams}, {Hancock}, {Lange}, {Tocknell},
		{Johnston-Hollitt}, {Kaplan}, {Tingay}, \& {Walker}}]{Bhat2023}
	{Bhat}, N.~D.~R., {Swainston}, N.~A., {McSweeney}, S.~J., {et~al.} 2023, \pasa,
	40, e021
	
	\bibitem[{{Brylyakova} \& {Tyul'bashev}(2021)}]{Brylyakova2021}
	{Brylyakova}, E.~A. \& {Tyul'bashev}, S.~A. 2021, \aap, 647, A191
	
	\bibitem[{{Burke-Spolaor} \& {Bailes}(2010)}]{Burke-Spolaor2010}
	{Burke-Spolaor}, S. \& {Bailes}, M. 2010, \mnras, 402, 855
	
	\bibitem[{{Burns} \& {Clark}(1969)}]{Burns1969}
	{Burns}, W.~R. \& {Clark}, B.~G. 1969, \aap, 2, 280
	
	\bibitem[{{Chen} \& {Ruderman}(1993)}]{Chen1993}
	{Chen}, K. \& {Ruderman}, M. 1993, \apj, 402, 264
	
	\bibitem[{{Cordes} \& {McLaughlin}(2003)}]{Cordes2003}
	{Cordes}, J.~M. \& {McLaughlin}, M.~A. 2003, \apj, 596, 1142
	
	\bibitem[{{Cui} {et~al.}(2017){Cui}, {Boyles}, {McLaughlin}, \&
		{Palliyaguru}}]{Cui2017}
	{Cui}, B.~Y., {Boyles}, J., {McLaughlin}, M.~A., \& {Palliyaguru}, N. 2017,
	\apj, 840, 5
	
	\bibitem[{{Deneva} {et~al.}(2013){Deneva}, {Stovall}, {McLaughlin}, {Bates},
		{Freire}, {Martinez}, {Jenet}, \& {Bagchi}}]{Deneva2013}
	{Deneva}, J.~S., {Stovall}, K., {McLaughlin}, M.~A., {et~al.} 2013, \apj, 775,
	51
	
	\bibitem[{{Dong} {et~al.}(2023){Dong}, {Crowter}, {Meyers}, {Pleunis},
		{Stairs}, {Tan}, {Yu}, {Boyle}, {Cook}, {Fonseca}, {Gaensler}, {Good},
		{Kaspi}, {McKee}, {Patel}, \& {Pearlman}}]{Dong2023}
	{Dong}, F.~A., {Crowter}, K., {Meyers}, B.~W., {et~al.} 2023, \mnras, 524, 5132
	
	\bibitem[{{Eatough} {et~al.}(2010){Eatough}, {Molkenthin}, {Kramer}, {Noutsos},
		{Keith}, {Stappers}, \& {Lyne}}]{Eatough2010}
	{Eatough}, R.~P., {Molkenthin}, N., {Kramer}, M., {et~al.} 2010, \mnras, 407,
	2443
	
	\bibitem[{{Fridman}(2010)}]{Fridman2010}
	{Fridman}, P.~A. 2010, \mnras, 409, 808
	
	\bibitem[{{Han} {et~al.}(2021){Han}, {Wang}, {Wang}, {Wang}, {Zhou}, {Sun},
		{Yan}, {Su}, {Jing}, {Chen}, {Gao}, {Hou}, {Xu}, {Lee}, {Wang}, {Jiang},
		{Xu}, {Yan}, {Gan}, {Guan}, {Huang}, {Jiang}, {Li}, {Men}, {Sun}, {Wang},
		{Wang}, {Wang}, {Xie}, {Xu}, {Yao}, {You}, {Yu}, {Yuan}, {Yuen}, {Zhang}, \&
		{Zhu}}]{Han2021}
	{Han}, J.~L., {Wang}, C., {Wang}, P.~F., {et~al.} 2021, Research in Astronomy
	and Astrophysics, 21, 107
	
	\bibitem[{{Hankins} \& {Rickett}(1975)}]{Hankins1975}
	{Hankins}, T.~H. \& {Rickett}, B.~J. 1975, Methods in Computational Physics,
	14, 55
	
	\bibitem[{{Hochreiter} \& {Schmidhuber}(1997)}]{Hochreiter1997}
	{Hochreiter}, S. \& {Schmidhuber}, J. 1997, {Long short-term memory}
	
	\bibitem[{{Kuz'min} {et~al.}(2007){Kuz'min}, {Losovskii}, \&
		{Lapaev}}]{Kuzmin2007}
	{Kuz'min}, A.~D., {Losovskii}, B.~Y., \& {Lapaev}, K.~A. 2007, Astronomy
	Reports, 51, 615
	
	\bibitem[{{Logvinenko} {et~al.}(2020){Logvinenko}, {Tyul'bashev}, \&
		{Malofeev}}]{Logvinenko2020}
	{Logvinenko}, S.~V., {Tyul'bashev}, S.~A., \& {Malofeev}, V.~M. 2020, Bulletin
	of the Lebedev Physics Institute, 47, 390
	
	\bibitem[{{Lorimer} \& {Kramer}(2012)}]{Lorimer2012}
	{Lorimer}, D.~R. \& {Kramer}, M. 2012, {Handbook of Pulsar Astronomy}
	
	\bibitem[{{Manchester} {et~al.}(2005){Manchester}, {Hobbs}, {Teoh}, \&
		{Hobbs}}]{Manchester2005}
	{Manchester}, R.~N., {Hobbs}, G.~B., {Teoh}, A., \& {Hobbs}, M. 2005, \aj, 129,
	1993
	
	\bibitem[{{Maron} {et~al.}(2000){Maron}, {Kijak}, {Kramer}, \&
		{Wielebinski}}]{Maron2000}
	{Maron}, O., {Kijak}, J., {Kramer}, M., \& {Wielebinski}, R. 2000, \aaps, 147,
	195
	
	\bibitem[{{McLaughlin} {et~al.}(2006){McLaughlin}, {Lyne}, {Lorimer}, {Kramer},
		{Faulkner}, {Manchester}, {Cordes}, {Camilo}, {Possenti}, {Stairs}, {Hobbs},
		{D'Amico}, {Burgay}, \& {O'Brien}}]{McLaughlin2006}
	{McLaughlin}, M.~A., {Lyne}, A.~G., {Lorimer}, D.~R., {et~al.} 2006, \nat, 439,
	817
	
	\bibitem[{{Sanidas} {et~al.}(2019){Sanidas}, {Cooper}, {Bassa}, {Hessels},
		{Kondratiev}, {Michilli}, {Stappers}, {Tan}, {van Leeuwen}, {Cerrigone},
		{Fallows}, {Iacobelli}, {Orr{\'u}}, {Pizzo}, {Shulevski}, {Toribio}, {ter
			Veen}, {Zucca}, {Bondonneau}, {Grie{\ss}meier}, {Karastergiou}, {Kramer}, \&
		{Sobey}}]{Sanidas2019}
	{Sanidas}, S., {Cooper}, S., {Bassa}, C.~G., {et~al.} 2019, \aap, 626, A104
	
	\bibitem[{{Shishov} {et~al.}(2016){Shishov}, {Chashei}, {Oreshko},
		{Logvinenko}, {Tyul'bashev}, {Subaev}, {Svidskii}, {Lapshin}, \&
		{Dagkesamanskii}}]{Shishov2016}
	{Shishov}, V.~I., {Chashei}, I.~V., {Oreshko}, V.~V., {et~al.} 2016, Astronomy
	Reports, 60, 1067
	
	\bibitem[{{Staelin}(1969)}]{Staelin1969}
	{Staelin}, D.~H. 1969, IEEE Proceedings, 57, 724
	
	\bibitem[{{Swinbank} {et~al.}(2015){Swinbank}, {Staley}, {Molenaar}, {Rol},
		{Rowlinson}, {Scheers}, {Spreeuw}, {Bell}, {Broderick}, {Carbone}, {Garsden},
		{van der Horst}, {Law}, {Wise}, {Breton}, {Cendes}, {Corbel},
		{Eisl{\"o}ffel}, {Falcke}, {Fender}, {Grie{\ss}meier}, {Hessels}, {Stappers},
		{Stewart}, {Wijers}, {Wijnands}, \& {Zarka}}]{Swinbank2015}
	{Swinbank}, J.~D., {Staley}, T.~D., {Molenaar}, G.~J., {et~al.} 2015, Astronomy
	and Computing, 11, 25
	
	\bibitem[{{Tingay} {et~al.}(2013){Tingay}, {Goeke}, {Bowman}, {Emrich}, {Ord},
		{Mitchell}, {Morales}, {Booler}, {Crosse}, {Wayth}, {Lonsdale}, {Tremblay},
		{Pallot}, {Colegate}, {Wicenec}, {Kudryavtseva}, {Arcus}, {Barnes},
		{Bernardi}, {Briggs}, {Burns}, {Bunton}, {Cappallo}, {Corey}, {Deshpande},
		{Desouza}, {Gaensler}, {Greenhill}, {Hall}, {Hazelton}, {Herne}, {Hewitt},
		{Johnston-Hollitt}, {Kaplan}, {Kasper}, {Kincaid}, {Koenig}, {Kratzenberg},
		{Lynch}, {Mckinley}, {Mcwhirter}, {Morgan}, {Oberoi}, {Pathikulangara},
		{Prabu}, {Remillard}, {Rogers}, {Roshi}, {Salah}, {Sault}, {Udaya-Shankar},
		{Schlagenhaufer}, {Srivani}, {Stevens}, {Subrahmanyan}, {Waterson},
		{Webster}, {Whitney}, {Williams}, {Williams}, \& {Wyithe}}]{Tingay2013}
	{Tingay}, S.~J., {Goeke}, R., {Bowman}, J.~D., {et~al.} 2013, \pasa, 30, e007
	
	\bibitem[{{Turtle} \& {Baldwin}(1962)}]{Turtle1962}
	{Turtle}, A.~J. \& {Baldwin}, J.~E. 1962, \mnras, 124, 459
	
	\bibitem[{{Tyul'bashev} {et~al.}(2023){Tyul'bashev}, {Kitaeva}, {Brylyakova},
		{Tyul'bashev}, \& {Tyul'basheva}}]{Tyulbashev2023}
	{Tyul'bashev}, S.~A., {Kitaeva}, M.~A., {Brylyakova}, E.~A., {Tyul'bashev},
	V.~S., \& {Tyul'basheva}, G.~E. 2023, Astronomy Letters, 49, 533
	
	\bibitem[{{Tyul'bashev} {et~al.}(2021{\natexlab{a}}){Tyul'bashev}, {Kitaeva},
		{Logvinenko}, \& {Tyul'basheva}}]{Tyulbashev2021a}
	{Tyul'bashev}, S.~A., {Kitaeva}, M.~A., {Logvinenko}, S.~V., \& {Tyul'basheva},
	G.~E. 2021{\natexlab{a}}, Astronomy Reports, 65, 1246
	
	\bibitem[{{Tyul'bashev} {et~al.}(2022{\natexlab{a}}){Tyul'bashev}, {Kitaeva},
		\& {Tyul'basheva}}]{Tyulbashev2022b}
	{Tyul'bashev}, S.~A., {Kitaeva}, M.~A., \& {Tyul'basheva}, G.~E.
	2022{\natexlab{a}}, \mnras, 517, 1112
	
	\bibitem[{{Tyul'bashev} {et~al.}(2022{\natexlab{b}}){Tyul'bashev}, {Pervukhin},
		{Kitaeva}, {Tyul'basheva}, {Brylyakova}, \& {Chernosov}}]{Tyulbashev2022a}
	{Tyul'bashev}, S.~A., {Pervukhin}, D.~V., {Kitaeva}, M.~A., {et~al.}
	2022{\natexlab{b}}, \aap, 664, A37
	
	\bibitem[{{Tyul'bashev} {et~al.}(2021{\natexlab{b}}){Tyul'bashev}, {Smirnova},
		{Brylyakova}, \& {Kitaeva}}]{Tyulbashev2021}
	{Tyul'bashev}, S.~A., {Smirnova}, T.~V., {Brylyakova}, E.~A., \& {Kitaeva},
	M.~A. 2021{\natexlab{b}}, \mnras, 508, 2815
	
	\bibitem[{{Tyul'bashev} {et~al.}(2017){Tyul'bashev}, {Tyul'bashev}, {Kitaeva},
		{Chernyshova}, {Malofeev}, {Chashei}, {Shishov}, {Dagkesamanskii},
		{Klimenko}, {Nikitin}, \& {Nikitina}}]{Tyulbashev2017}
	{Tyul'bashev}, S.~A., {Tyul'bashev}, V.~S., {Kitaeva}, M.~A., {et~al.} 2017,
	Astronomy Reports, 61, 848
	
	\bibitem[{{Tyul'bashev} {et~al.}(2018{\natexlab{a}}){Tyul'bashev},
		{Tyul'bashev}, \& {Malofeev}}]{Tyulbashev2018b}
	{Tyul'bashev}, S.~A., {Tyul'bashev}, V.~S., \& {Malofeev}, V.~M.
	2018{\natexlab{a}}, \aap, 618, A70
	
	\bibitem[{{Tyul'bashev} {et~al.}(2018{\natexlab{b}}){Tyul'bashev},
		{Tyul'bashev}, {Malofeev}, {Logvinenko}, {Oreshko}, {Dagkesamanskii},
		{Chashei}, {Shishov}, \& {Bursov}}]{Tyulbashev2018a}
	{Tyul'bashev}, S.~A., {Tyul'bashev}, V.~S., {Malofeev}, V.~M., {et~al.}
	2018{\natexlab{b}}, Astronomy Reports, 62, 63
	
	\bibitem[{Tyul'bashev {et~al.}(2016)Tyul'bashev, Tyul'bashev, Oreshko, \&
		Logvinenko}]{Tyulbashev2016}
	Tyul'bashev, S.~A., Tyul'bashev, V.~S., Oreshko, V.~V., \& Logvinenko, S.~V.
	2016, Astronomy Reports, 60, 220
	
	\bibitem[{{Tyul'bashev} {et~al.}(2024){Tyul'bashev}, {Tyul'basheva}, {Kitaeva},
		{Ovchinnikov}, {Oreshko}, \& {Logvinenko}}]{Tyulbashev2024}
	{Tyul'bashev}, S.~A., {Tyul'basheva}, G.~E., {Kitaeva}, M.~A., {et~al.} 2024,
	\mnras, 528, 2220
	
	\bibitem[{{van Haarlem} {et~al.}(2013){van Haarlem}, {Wise}, {Gunst}, {Heald},
		{McKean}, {Hessels}, {de Bruyn}, {Nijboer}, {Swinbank}, {Fallows},
		{Brentjens}, {Nelles}, {Beck}, {Falcke}, {Fender}, {H{\"o}randel},
		{Koopmans}, {Mann}, {Miley}, {R{\"o}ttgering}, {Stappers}, {Wijers},
		{Zaroubi}, {van den Akker}, {Alexov}, {Anderson}, {Anderson}, {van Ardenne},
		{Arts}, {Asgekar}, {Avruch}, {Batejat}, {B{\"a}hren}, {Bell}, {Bell}, {van
			Bemmel}, {Bennema}, {Bentum}, {Bernardi}, {Best}, {B{\^\i}rzan}, {Bonafede},
		{Boonstra}, {Braun}, {Bregman}, {Breitling}, {van de Brink}, {Broderick},
		{Broekema}, {Brouw}, {Br{\"u}ggen}, {Butcher}, {van Cappellen}, {Ciardi},
		{Coenen}, {Conway}, {Coolen}, {Corstanje}, {Damstra}, {Davies}, {Deller},
		{Dettmar}, {van Diepen}, {Dijkstra}, {Donker}, {Doorduin}, {Dromer}, {Drost},
		{van Duin}, {Eisl{\"o}ffel}, {van Enst}, {Ferrari}, {Frieswijk}, {Gankema},
		{Garrett}, {de Gasperin}, {Gerbers}, {de Geus}, {Grie{\ss}meier}, {Grit},
		{Gruppen}, {Hamaker}, {Hassall}, {Hoeft}, {Holties}, {Horneffer}, {van der
			Horst}, {van Houwelingen}, {Huijgen}, {Iacobelli}, {Intema}, {Jackson},
		{Jelic}, {de Jong}, {Juette}, {Kant}, {Karastergiou}, {Koers}, {Kollen},
		{Kondratiev}, {Kooistra}, {Koopman}, {Koster}, {Kuniyoshi}, {Kramer},
		{Kuper}, {Lambropoulos}, {Law}, {van Leeuwen}, {Lemaitre}, {Loose}, {Maat},
		{Macario}, {Markoff}, {Masters}, {McFadden}, {McKay-Bukowski}, {Meijering},
		{Meulman}, {Mevius}, {Middelberg}, {Millenaar}, {Miller-Jones}, {Mohan},
		{Mol}, {Morawietz}, {Morganti}, {Mulcahy}, {Mulder}, {Munk}, {Nieuwenhuis},
		{van Nieuwpoort}, {Noordam}, {Norden}, {Noutsos}, {Offringa}, {Olofsson},
		{Omar}, {Orr{\'u}}, {Overeem}, {Paas}, {Pandey-Pommier}, {Pandey}, {Pizzo},
		{Polatidis}, {Rafferty}, {Rawlings}, {Reich}, {de Reijer}, {Reitsma},
		{Renting}, {Riemers}, {Rol}, {Romein}, {Roosjen}, {Ruiter}, {Scaife}, {van
			der Schaaf}, {Scheers}, {Schellart}, {Schoenmakers}, {Schoonderbeek},
		{Serylak}, {Shulevski}, {Sluman}, {Smirnov}, {Sobey}, {Spreeuw}, {Steinmetz},
		{Sterks}, {Stiepel}, {Stuurwold}, {Tagger}, {Tang}, {Tasse}, {Thomas},
		{Thoudam}, {Toribio}, {van der Tol}, {Usov}, {van Veelen}, {van der Veen},
		{ter Veen}, {Verbiest}, {Vermeulen}, {Vermaas}, {Vocks}, {Vogt}, {de Vos},
		{van der Wal}, {van Weeren}, {Weggemans}, {Weltevrede}, {White}, {Wijnholds},
		{Wilhelmsson}, {Wucknitz}, {Yatawatta}, {Zarka}, {Zensus}, \& {van
			Zwieten}}]{vanHaarlem2013}
	{van Haarlem}, M.~P., {Wise}, M.~W., {Gunst}, A.~W., {et~al.} 2013, \aap, 556,
	A2
	
	\bibitem[{{Wang} {et~al.}(2007){Wang}, {Manchester}, \& {Johnston}}]{Wang2007}
	{Wang}, N., {Manchester}, R.~N., \& {Johnston}, S. 2007, \mnras, 377, 1383
	
	\bibitem[{{Weltevrede} {et~al.}(2006){Weltevrede}, {Stappers}, {Rankin}, \&
		{Wright}}]{Weltevrede2006}
	{Weltevrede}, P., {Stappers}, B.~W., {Rankin}, J.~M., \& {Wright}, G.~A.~E.
	2006, \apjl, 645, L149
	
	\bibitem[{{Yao} {et~al.}(2017){Yao}, {Manchester}, \& {Wang}}]{Yao2017}
	{Yao}, J.~M., {Manchester}, R.~N., \& {Wang}, N. 2017, \apj, 835, 29
	
	\bibitem[{{Zhang} {et~al.}(2007){Zhang}, {Gil}, \& {Dyks}}]{Zhang2007}
	{Zhang}, B., {Gil}, J., \& {Dyks}, J. 2007, \mnras, 374, 1103
	
	\bibitem[{{Zhang} {et~al.}(2020){Zhang}, {Wang}, {Hobbs}, {Russell}, {Li},
		{Zhang}, {Dai}, {Wu}, {Pan}, {Zhu}, {Toomey}, \& {Ren}}]{Zhang2020}
	{Zhang}, C., {Wang}, C., {Hobbs}, G., {et~al.} 2020, \aap, 642, A26
	
	\bibitem[{{Zhou} {et~al.}(2023){Zhou}, {Han}, {Xu}, {Wang}, {Wang}, {Wang},
		{Jing}, {Chen}, {Yan}, {Su}, {Gan}, {Jiang}, {Sun}, {Wang}, {Wang}, {Wang},
		{Xu}, \& {You}}]{Zhou2023}
	{Zhou}, D.~J., {Han}, J.~L., {Xu}, J., {et~al.} 2023, Research in Astronomy and
	Astrophysics, 23, 104001
	
\end{thebibliography}

\end{document}